\documentclass[fleqn,usenatbib]{mnras}

\usepackage{newtxtext,newtxmath}

\usepackage[T1]{fontenc}

\DeclareRobustCommand{\VAN}[3]{#2}
\let\VANthebibliography\thebibliography
\def\thebibliography{\DeclareRobustCommand{\VAN}[3]{##3}\VANthebibliography}

\usepackage[utf8]{inputenc}
\usepackage{bm,amsmath}  
\usepackage{natbib}
\usepackage{aas_macros}
\usepackage{caption,subcaption,setspace,graphicx,color}
\usepackage{booktabs,array}
\usepackage{multirow,makecell}  
\graphicspath{{figures/}}
\usepackage{xcolor}

 \setcitestyle{notesep={}}

\title[Asteroid Magnetization from the Early Solar Wind]{Asteroid Magnetization from the Early Solar Wind}

\author[A. Anand et al.]{Atma Anand$^{1}$\thanks{aanand6@ur.rochester.edu}, Jonathan Carroll-Nellenback$^{1,3}$, Eric G.~Blackman$^{1,3}$, John A.~Tarduno$^{1,2}$ \\
$^{1}$Department of Physics and Astronomy, University of Rochester, Rochester NY 14627\\
$^{2}$Department of Earth and Environmental Sciences, University of Rochester, Rochester NY 14627 \\
$^{3}$Laboratory for Laser Energetics, University of Rochester, Rochester, NY 14623, USA \\}

\date{September 2021}

\pubyear{2021}
    
\DeclareRobustCommand{\uppartial}{\text{\rotatebox[origin=t]{16}{\scalebox{0.95}[1]{$\partial$}}}\hspace{-1pt}}

\begin{document}
\label{firstpage}
\pagerange{\pageref{firstpage}--\pageref{lastpage}}
\maketitle

\begin{abstract}
Magnetic fields provide an important probe of  the thermal, material, and structural history of  planetary and sub-planetary bodies. 
Core dynamos are a potential source of magnetic fields for differentiated bodies, but evidence of  magnetization in undifferentiated bodies requires a different mechanism. 
Here we study the amplified field provided by the stellar wind to an initially unmagnetized body using analytic  theory and numerical simulations, employing the resistive MHD  \textsc{AstroBEAR} adaptive mesh refinement (AMR) multiphysics code.  We  obtain a broadly applicable scaling relation for the peak magnetization achieved once a wind advects, piles-up, and drapes a body with magnetic field,  reaching a quasi-steady state. We find that  the dayside magnetic field for a sufficiently conductive body saturates when it balances the sum of incoming solar wind ram, magnetic, and thermal pressures. Stronger amplification results from  pileup by denser and faster  winds.  Careful quantification of numerical diffusivity is required for accurately interpreting the peak magnetic field strength from  simulations, and corroborating with theory. As specifically applied to the Solar System, we find that early solar wind-induced  field amplification is a viable source of magnetization  
for observed paleointensities in meteorites from some undifferentiated bodies.
This  mechanism may also be applicable to other Solar System bodies, including metal-rich bodies to be visited in future space missions such as the asteroid (16) Psyche.

\end{abstract}




\section{Introduction}
Core dynamos in rocky bodies are possible when such objects have undergone differentiation and have sufficient energy to sustain connection in a liquid metal core. Mercury \citep[e.g.][]{cJohnson2015} and Mars \citep[e.g.][]{acuna1999} have each at some time hosted internal convective core dynamos. 
Earth appears to have had a core dynamo for nearly all of its history, before and after growth of the solid inner core \citep[see][; and references therein]{Olson2013,Bono2019,Tarduno2020}. 
The Moon’s small core was once thought to have hosted a dynamo \citep[e.g.][]{Cisowski1983}. However, recent analyses suggest that the Moon may have never had a long-lived field core dynamo, although a very early internally-generated field remains a possibility \citep{Tarduno2021}. 
It is often unclear whether differentiation generating a liquid core occurred in smaller rocky Solar System bodies like asteroids. If they are magnetized but undifferentiated, then their magnetic fields would have to be externally supplied.

Information about early asteroids - especially those destroyed in the collisions ubiquitous to development of the Solar System - comes from the study of meteorites. In general, paleofield strength data from paleomagnetic measurements can be used to detect the ambient magnetic  field during the formation of magnetic minerals in the studied meteorites. These potential field sources include those of  the protoplanetary disk nebula \citep{nagata1979}, solar wind, or dynamos in the parent bodies of meteorite parent bodies \citep{collinson1994}.

Some achrondite meteorites are clearly from highly differentiated parent bodies, and paleomagnetic measurements from multiple groups have revealed evidence for strong parent body paleofields.  For example, pallasites--meteorites that are composed of olivine and FeNi metals-- are clearly derived from a highly differentiated asteroid. High paleofields from the main group pallasites were first detected by magnetic studies of olivine bearing magnetic inclusions; these data indicated the presence of a core dynamo after an impact that formed the characteristic pallasite textures \citep{tarduno2012}. Dynamo sourced paleofields were also reported from magnetic studies of nanostructures in the FeNi metal \citep{bryson2014a, bryson2014b, nichols2021}. In contrast, meteorites from undifferentiated parent bodies with magnetic minerals formed after dispersal of the protoplanetary disk nebula, are expected to record only magnetic fields supplied by the solar wind. 

In this context, an important case is the Allende meteorite - 
a CV (Vigrano type) carbonaceous chondrite, long thought to be a classic example of a primitive meteorite from an undifferentiated body \citep{scott2014}. Strong apparent magnetizations were  known for Allende from classic studies  (e.g. \citealp{nagata1979}) but only more recently were these interpreted as requiring  paleofields 10's of $ \mathrm{\mu T} $, de facto evidence for a core dynamo, and  differentiation of the CV type parent asteroid \citep{Carporzen2011,weiss2013,elkins2011}. But this interpretation of the magnetic data has been the only evidence of CV parent body differentiation.
There is no evidence for differentiated components in meteorite collections.

The question of differentiation of the CV parent body was re-investigated by \cite{obrein2016lpsc} who discovered that Allende's magnetization was a  consequence of magnetic interactions associated with its magnetic mineralogy - especially pyrrhotite - rather than a reliable paleofield measurement. Allende's magnetization is thus not  useful for determining early ambient Solar System magnetic fields. 
A full account of the rock magnetic basis for Allende’s extreme magnetic interactions is provided in \cite{obrein20} (see in particular Supplementary Information Sections 1-6).
While this offered a solution to the Allende magnetization puzzle, indicating that this meteorite came from an undifferentiated (CV) asteroid as suggested by decades of prior work \citep[see][]{scott2014}, the question of what early ambient field strengths could arise has remained.
 
Fortunately, other CV as well as CM (Mighei-type) carbonaceous chondrites meteorites have more reliable magnetic recorders (i.e., magnetite) that can  preserve  useful indicators and constrain early Solar System magnetic fields. 
These yield paleofield strength values $\mathcal{O} \sim 1 \mu$T \citep{Cournede2015}.
\cite{tarduno2017lpi} proposed solar wind induced magnetization (WIM) as a plausible mechanism to explain the magnetization of these meteorite parent bodies, and outlined theory supporting this interpretation. \cite{oran18} produced a series of magnetohydrodynamic (MHD) simulations and argued that the solar wind was an insufficient source. However,  \cite{obrein20}  showed that with early solar wind parameters within an accepted range of observations and current theory \citep{a26,a27,Wood+2005,TARDUNO2014}, a solar system object at $ \sim 2-4 $ AU having a sufficiently conductive parent body or a thin conductive shell  may pile up sufficient magnetic field strength to explain the paleointensity data from CV, and CM meteorites having reliable magnetic recorders.
To see how the amplified solar field magnetization relates to meteorite magnetization values, see Figure 3 of \cite{obrein20}.

The solar wind MHD discussion in \cite{obrein20} and the detailed study of  the present paper focus on
the ambient external fields for near surface rocks on asteroids. The actual magnetization of these rocks will be further dependent on the nature of magnetic minerals present, when these minerals passed through their blocking temperatures (for thermoremanent magnetizations, TRM) or when they grew through their blocking volumes (in the case of chemical remanent magnetization, CRM). In addition, the time elapsed when passing through these unblocking temperatures or volumes is important.
In general, short TRM or CRM acquisition times will favor unidirectional magnetizations being imparted to near surface rocks on asteroids given reversals of the solar wind field. These factors are considered in detail in \cite{obrein20} (see Supplementary Information therein, sections 7-8).
CV and CM meteorite examples are discussed that have radiometric ages constraining magnetite formation to times after dispersal of nebular gas and dust, at 4.8 and 4.2 million years after CAI formation, respectively \citep{Fujiya2012, Doyle2015}.
The CM age assignment is based on the clustered Mn-Cr radiometric ages of CM carbonates. An older, precise magnetization age (2.9 $ \pm $ 0.39 Ma after CAIs) reported by \cite{Weiss2021},
based on their recalibration of I-Xe ages on magnetite reported in 
\cite{Pravdivtseva2013},
is not supported by the CM Mn-Cr ages, or a consideration of hydrothermal processes that can lead to the formation of magnetite and chemical remanent magnetization at relatively low temperatures \citep[see Section 7,][]{obrein20}. 
Mn-Cr ages of CM calcites \citep[measured with matrix-matched standard;][]{Fujiya2012} are based on the internal isochrons, and, therefore, are more reliable than I-Xe ages (A.N. Krot, personal communication). 
The older age is also inconsistent with the oldest bound ($\sim$ 3.2 Ma after CAIs) on the accretion age of the CM parent body discussed in \cite{Fujiya2012}.

Textural and magnetic unblocking data indicate rapid magnetic mineral formation in select CV and CM meteorites (see \cite{obrein20}, Supplementary Section 7). While these observations are consistent with an external magnetization provided by the amplified solar wind, 
they do not necessarily universally apply.
Other meteorites need to be evaluated on a case-by-case basis, considering the nature and formation time of their magnetic minerals.

In this manuscript, we carry out a more general numerical study of the ambient magnetic fields created on small bodies by the solar wind. We obtain broadly applicable scaling relations. These are especially relevant for early Solar System bodies known from meteorites, or to be visited by future space missions such as that to investigate (16) Psyche \citep{Elkins-Tanton+2020}, that could have near surface rocks with magnetic minerals formed 
after dispersal of nebular gas and dust.
In \S2, we discuss our numerical model and methods used, in \S3 we provide and discuss theoretical estimates, in \S4 we showcase our results, in \S5 we interpret  our findings and identify directions for  future work, and in \S6 we conclude our results.

\section{Model and Methods}\label{sec:model}

Using the adaptive mesh-refinement (AMR) code, AstroBEAR \citep{cunningham09,carroll13}, we simulate a magnetized solar wind overrunning an asteroid. For different runs we use different resistivity values and profiles for the asteroid, and each run was evolved to a steady state.

To carry out our simulations, we need a model for the early solar wind, the resistivity profile of the asteroid, initial and boundary conditions, and criteria for a sufficiently steady-state. We can then compare the theoretical framework of  \S\ref{sec:theory} to  the numerical results.

\subsection{Solar wind and grid setup}
The early solar wind is thought to have a higher density and magnetic field strength  than that of the present day, based on observations and modeling of young stars similar to the Sun \citep{Wood+2005,TARDUNO2014,a26,a27}. We give the young stellar wind an ion density of 1000 $\mbox{cm}^{-3}$ or 300 $\mbox{cm}^{-3}$, a temperature of  $ 5 \times 10^5$ K, a velocity of 500 $ \mathrm{km s^{-1}} $ (in $x$), and  a  magnetic field of 100 nT (in the $ y $ direction) perpendicular to the flow.  The above values correspond to a stellar wind Mach number of 4.74, 
the ratio of thermal to magnetic pressure $\beta$ of 3.47 or 1.04, and an Alfv\'enic Mach number of 8 or 4.38. We use an separate wind having parameters of 300 $\mbox{cm}^{-3}$, $ 150 \times 10^3$ K, 500 $ \mathrm{km \, s^{-1}} $, and 14 nT to test the effects of our grid (see Table \ref{tab:summary} for more details). The mean mass of solar wind particles in all cases is 0.623 amu, considering that Hydrogen, and Helium atoms will be ionized. 

The asteroid is modeled as a solid boundary of radius 500 km but allowed to have different interior  resistivity structures.  The solution results can be  rescaled  to arbitrary asteroid radii,  provided that the asteroid radius exceeds the ion Larmor radius. The wind density, temperature, velocity, and magnetization can also be rescaled provided that the same plasma $\beta$ and Mach number are used. 

We initialize the grid with the aforementioned stellar wind conditions, but make modifications in the vicinity of the asteroid as described below. The wind is  continuously injected from the left $(-x)$ boundary during the simulation.  The $y$ and $z$
simulation boundaries are 
extrapolated  as is the right boundary $(+x)$, allowing for the wind, and the wake behind the asteroid, to flow through the boundary. The boundaries are sufficiently far away such that the wake behind the asteroid does not pass through them - and the solar wind remains unperturbed (plane parallel).

We use a base resolution of 64x64x64 with 4 additional levels of adaptive mesh refinement (AMR) around the asteroid, allowing resolution of the asteroid's diameter with 128 zones. We performed a resolution study, comparing runs from 3 to 5 AMR levels which confirmed that the simulated field is sufficiently resolved  at the asteroid boundary, and a convergence study which confirmed that  the field reached steady state for all models (details are excluded from this study).

\subsection{Equations solved}
Simulations were conducted using AstroBEAR \footnote{https://astrobear.pas.rochester.edu/} \citep{cunningham09,carroll13}, a publicly available, massively parallelized, adaptive mesh refinement (AMR) code that contains a variety of multi-physics solvers (i.e. self-gravity, magnetic resistivity, radiative transport, ionization dynamics, heat conduction, etc).  The physical mass, momentum and energy equations solved are 
\begin{equation}
    \frac{\uppartial \rho}{\uppartial t} + \boldsymbol{\nabla} \cdot \left ( \rho \boldsymbol{v} \right ) = 0 
    \label{eq:Eu1}
\end{equation}
\begin{equation}
    \frac{\uppartial \rho \boldsymbol{v}}{\uppartial  t} + \boldsymbol{\nabla} \cdot \left ( \rho \boldsymbol{v} \boldsymbol{v}  - \boldsymbol{BB} \right )= - \boldsymbol{\nabla} P^*
    \label{eq:Eu2}
\end{equation}
\begin{equation}
    \frac{\uppartial  E}{\uppartial  t} + \boldsymbol{\nabla} \cdot \left [ (E + P^*) \boldsymbol{v} - \boldsymbol{B} \left (\boldsymbol{B} \cdot \boldsymbol{v} \right) \right ] =  - \boldsymbol{\nabla} \cdot \left ( \eta \boldsymbol{J}  \times \boldsymbol{B}\right )
    \label{eq:Eu3}
\end{equation}

\begin{equation}
    \frac{\uppartial  \boldsymbol{B}}{\uppartial  t}- \boldsymbol{\nabla} \times \left ( \boldsymbol{v} \times \boldsymbol{B} \right ) =-\boldsymbol{\nabla} \times (\eta \boldsymbol{J}) 
    \label{eq:Eu4}
\end{equation}
where $\rho$ is the mass density, $\boldsymbol{v}$ is the velocity, $P^* = P + \frac{B^2}{2}$ where $P$ is the gas pressure, $\boldsymbol{B}$ is the magnetic field,  the total energy $E = \frac{P}{\gamma - 1} + \frac{1}{2}\rho v^2 + \frac{B^2}{2}$ and the magnetic diffusivity is given by $\eta$.  We set $\gamma = \frac{5}{3}$ and use the non-relativistic approximation $ \boldsymbol{J} = \boldsymbol{\nabla} \times \boldsymbol{B} $.

In practice, we found it necessary to add a small amount of explicit numerical diffusion throughout the simulation domain (similar to that used in higher order schemes to damp oscillations at regions of flow convergence) to smooth over the discrete representation of the asteroid surface on our Cartesian grid.  This was accomplished by adding $-\alpha \boldsymbol{\nabla} \boldsymbol{U}$ to the corresponding fluxes for the conserved fluid variables $\boldsymbol{U} = \left [ \rho, \rho \boldsymbol{v}, E \right]$.  

\begin{equation}
\frac{\uppartial  \boldsymbol{U}}{\uppartial  t} = \boldsymbol{\nabla} \cdot \left( \alpha \boldsymbol{\nabla} \boldsymbol{U} \right)
\end{equation}

\noindent where $\alpha$ was chosen to be as small as possible.  In addition the magnetic field was updated in a divergence conserving manner using

\begin{equation}
    \frac{\uppartial  \boldsymbol{B}}{\uppartial  t} = - \boldsymbol{\nabla} \times \left (\alpha \boldsymbol{\nabla} \times \boldsymbol{B} \right ) = -\boldsymbol{\nabla} \times ( \alpha \boldsymbol{J} ),
\end{equation}

\noindent so that the right side  acted like a source of additional magnetic diffusion throughout the simulation domain. As the simulation approached steady state, the 
$\alpha$ was  ramped down to values small enough to  smooth out gradients only on the finest grid scale - and to levels where the additional diffusion was comparable to the intrinsic numerical diffusion.  This additional diffusion was only applied outside the asteroid. 

\subsection{Initial velocity and magnetic fields}
Asteroids could potentially have outer layers of higher conductivity or atmospheres, particular in the young Solar System \citep{brownlee,obrein20}, and so a number of our runs focus on asteroid surfaces with  low  resistivity so that  the initial magnetic field takes a long time to diffuse through and  reach a steady state. Hence, we initialize the magnetic field to exclude the asteroid and a couple of zones outside it. To enforce $\boldsymbol{\nabla}\cdot \boldsymbol{B}=0$, we solve for the magnetic vector potential $\boldsymbol{A}$ (where $\boldsymbol{B}=\boldsymbol{\nabla} \times \boldsymbol{A}$) of a magnetically shielded sphere with no field inside and a fixed $ B_0 $ field at infinity. We have
\begin{equation} \label{eq:binit}
    \boldsymbol{A} = 
    \begin{cases}
        B_0 \left(1-\frac{r_e^3}{r^3}\right) \boldsymbol{\hat{B}} \times \boldsymbol{r} \, & ; \ r > r_e \\
        0 \, & ; \ r \le r_e \\ 
     \end{cases}
\end{equation}
\noindent  where $ B_0 $ is taken to be that of the solar wind,  $r_e$ is the exclusion radius of the magnetic field, $\boldsymbol{r}$ is the position vector from the center of the asteroid, and $ \boldsymbol{\hat{B}}$ is the direction of the magnetic field. The assumed incompressible  wind velocity vector is chosen to be proportional to the magnetic field. That is, the curl of Eqn. \ref{eq:binit} was taken, transposed, and scaled with a peak value equal to that of the solar wind velocity. This also makes it divergence free, as one would expect for an incompressible flow around a spherical obstruction.

\begin{figure}
	\begin{center}
		\includegraphics[width=0.48 \textwidth]{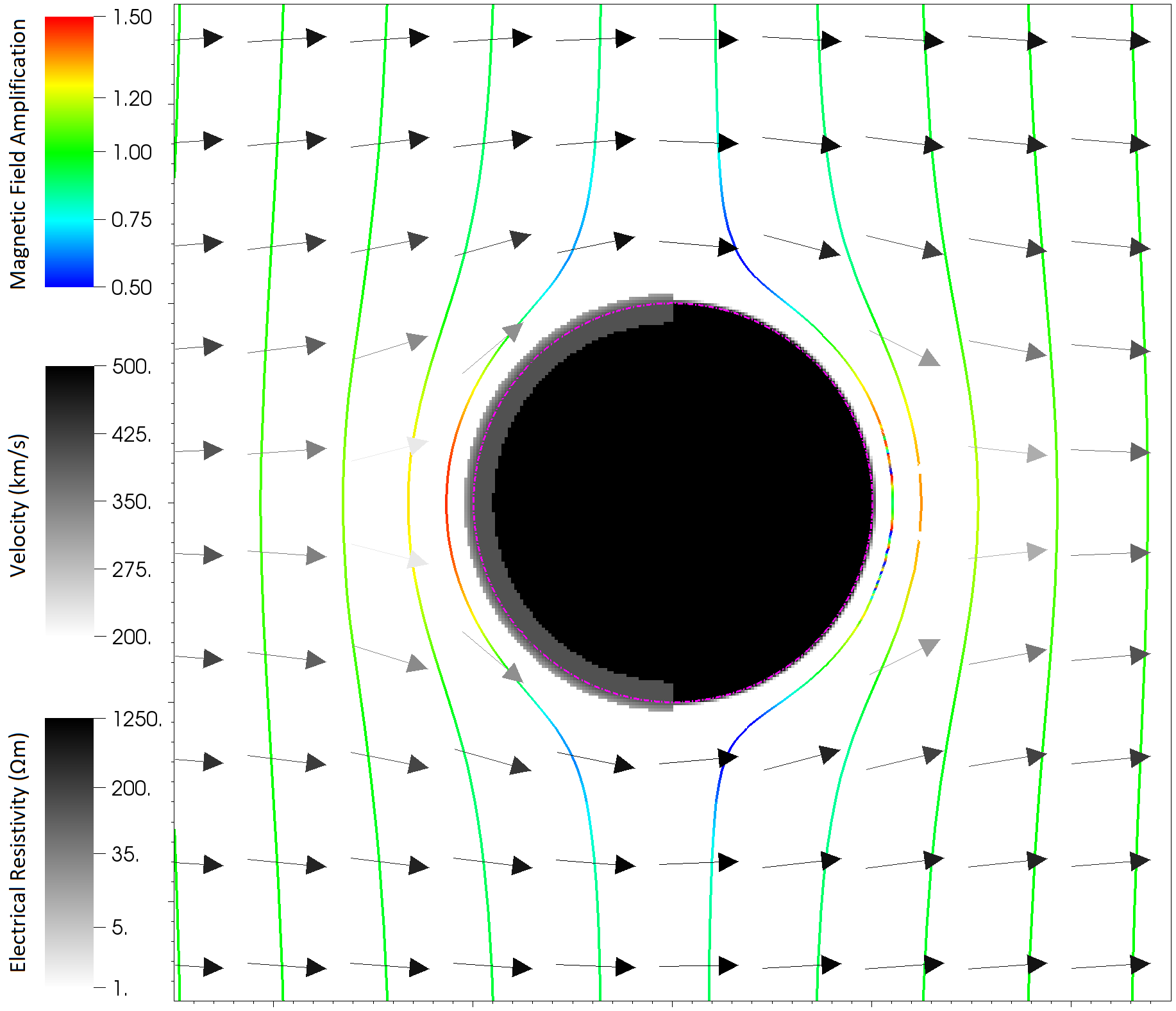}
		\caption{Initial values of the magnetic field amplification (lines), and velocity (arrows) as per Eqn. \ref{eq:binit}, and comet-like resistivity structure in the xy mid-plane. The resistivity of the ambient is negligible. We also use uniform (fully black colored body) and shell-like resistivity (grey shell extends to both hemispheres) profiles. There is an exponential resistivity smoothing of 3 zones merging into the ambient. The particle density inside the body is fixed to be low ($ \sim 10$ ions), while the ambient wind has density of $ 1000 \ \mbox{cm}^{-3} $ and field of $ 100\ \text{nT} $.}
		\label{fig:init}
	\end{center}
\end{figure}

Fig. \ref{fig:init} shows a slice in the $xy$ midplane of the solar wind velocity, electrical resistivity, and ambient magnetic field setup for the initial ($t=0$) frame. The wind velocity is in the $x$ direction, carrying magnetic field in the $y$ direction. Note the exclusion of the field and wind from the asteroid, and the similar setup for the  wind velocity and magnetic field.

\subsection{Asteroid boundary conditions}
The asteroid is incorporated into the simulation box as an internal partially absorbing boundary similar to \cite{jia15}.  When the flow is into the boundary, the flow density and internal energy are extrapolated
and the flow velocity is multiplied by an absorption factor to allow for control of how much mass, momentum, and energy is absorbed.  When this factor is 1, the boundary fully absorbs the incoming mass, momentum, and energy. When the factor is -1, the boundary reflects, allowing no mass, or energy to pass through. We find that the value of the absorption factor makes observable differences in the final steady state solution -- particularly for high magnetic Reynolds' number ($R_M$) cases, with absorbing boundaries having more field pileup.

When the flow is away from the boundary, the boundary acts as a  low density and pressure region, and  allows  some material to be drawn out through it. On the night side, this extra material helps to limit the Alfv\'en speeds that could otherwise be produced by the diffusive flux of magnetic fields through  this boundary into what would otherwise be a very low density environment. 
The initial asteroid density is also  chosen high enough to avoid extremely high Alfv\'en speeds  behind the asteroid, but much lower than the ambient wind so as to not be a significant source of particles.

The asteroid density is slowly ramped down over the course of the simulations to limit its influence on the exterior magnetic structure and facilitate a steady state.  In all cases, there is no advection of magnetic fields across the boundary or within the asteroid,  only magnetic diffusion.

\subsection{Asteroid interior resistivity}\label{sec:res}
We model our asteroid as a body with radius $R=500 \, \text{km}$ and  three different resistivity profiles.
\begin{enumerate}
	\item \textbf{Uniform:} In this case, the resistivity is constant with radius. This is the simplest case and can be readily  extended to a family of models with different resistivities, all other conditions being the same. This case corresponds to  cool, small asteroids or moons, that are low in volatiles.
	
	\item \textbf{Conductive Outer Shell:} For this case the outer 10\% shell is 10 times more conductive than the body. This allows the field to pile up more effectively at the outside boundary of the shell. The shell could correspond to a surface layer of enhanced conductivity within the body itself, or an atmospheric layer above the body.  Venus, with its dense atmosphere would be an extreme example of  the latter.
	
	\item \textbf{Cometary:} In this case the outer 10\% shell on the dayside has a higher conductivity compared to that on the night side and  body of the asteroid, the latter two of which are equal (see Fig. \ref{fig:init}). This case is similar the interaction of comets with the solar wind, which leads to volatiles escaping and forming a conductive layer on the dayside. Such an interaction may be particulaly relevant  for the early solar wind \citep{brownlee,obrein20}. Fig. \ref{fig:init} exemplifies this case for a body of $ R_M = 500 $ and dayside shell of $ R_M = $ 5,000.
\end{enumerate}
The total size of the object is the same in all cases. The shell or cometary layer is included in the simulation at the expense of  zones in the body.

For the cases with uniform diffusivity, $\eta$, the magnetic Reynolds number $R_M \equiv \frac{2 v_w R}{\eta}$ and can be thought of as the ratio of advective flux around the asteroid $\left (v_w B \right)$ to diffusive flux through the asteroid $\left (\frac{\eta B}{2 R}\right)$.  
For models with $R_M \lesssim 1$, magnetic fields  readily diffuse through the asteroid, while for models with $R_M >> 1$, magnetic fields  pile up on the upwind side faster than they can diffuse through the body. The latter leads to  an preferential increase  of magnetic pressure over thermal pressure because material can  escape along field lines parallel to the body, leaving the magnetic pressure to balance and divert the the solar wind around the asteroid. For the cases with non-uniform diffusivity, we compute an the magnetic Reynolds using an effective diffusivity $\bar{\eta}$ which we define as the inverse of the current weighted conductivity 

\begin{equation}\label{eq:eqR}
	\frac{1}{\bar{\eta}} = \bar{\sigma} = \frac{\displaystyle \int{ \  \sigma  \ \left | \boldsymbol{\nabla}\times \boldsymbol{B} \right | \ \text{d} V}}{\displaystyle \int{  \ \left | \boldsymbol{\nabla}\times \boldsymbol{B} \right | \ \text{d} V}} = \frac{\displaystyle \int{ \  \frac{1}{\eta}  \ \left | \boldsymbol{\nabla}\times \boldsymbol{B} \right | \ \text{d} V}}{\displaystyle \int{  \ \left | \boldsymbol{\nabla}\times \boldsymbol{B} \right | \ \text{d} V}}
\end{equation}

For our choice of solar wind speed $v_w=500 \, \text{km s}^{-1}$ and asteroid radius of 500 km, the critical magnetic diffusivity $\eta_c$ for which $R_M = 1$ is $5 \times 10^{11} \, \text{m}^2 \text{s}^{-1}$,  corresponding to an electrical resistivity $\approx 6.3 \times 10^5 \, \Omega \text{m}$.

For the  asteroid electrical resistivity ($ \rho_e $) we use values based on the heating curve of Sample 1 of Allende (heated at $ 3 ^{\circ} \text{C hr}^{-1} $) in \cite{duba84} at $ 300 ^{\circ} \text{C} $, consistent with not having a differentiated interior. This gives $\rho_e \sim 1250 \, \Omega m $, and a magnetic Reynolds number, $R_M \approx 500$.

\subsection{Steady state criteria}
Wherever possible, all simulations were run to a maximum of $ \sim 72 $ seconds 
or a suitable fraction of the diffusion time periods. The steady state was verified by verifying that the rate of change of maximum field drops to a negligible value  in successive output frames. 

\section{Limits from analytic theory} 
\label{sec:theory}

\subsection{Limits on magnetic amplification due to asteroid diffusivity}
If we consider a steady state solution in 1D, the induction equation becomes
\begin{equation}
    \frac{d}{\text{d} x} \left ( v B - \eta \frac{\text{d} B}{\text{d} x} \right ) = 0
\end{equation}
which states that the combined advective  diffusive flux of the magnetic field must be constant and equal to the incoming advective flux of the solar wind. 
\begin{equation}
    v B - \eta \frac{\text{d} B}{\text{d} x} = v_w B_0.
    \label{7}
\end{equation}
Within any slab of width $2 R$ interior to an asteroid, $v = 0$ and equation (\ref{7}) becomes
\begin{equation} 
    \eta \frac{\text{d}B}{\text{d} x} = v_w B_0
\end{equation}

If we further assume that any fields which diffuse through the asteroid from left to right are quickly removed,   so that the field at the right edge of the slab  $\approx 0$, the  field at the inflowing edge needed to provide a sufficient gradient to diffuse the field through the body as quickly as it is being advected in can be estimated as
\begin{equation}
    B_{\max} = v_w B_0 \displaystyle \int_0^{2 R}{\frac{\text{d} x}{\eta}} = B_0 \frac{2 v_w R}{\bar{\eta}} = B_0 R_M, 
\end{equation}
and the expected amplification factor 
\begin{equation}
    A \equiv \frac{B_{\max}}{B_0} =  R_M
\end{equation}
where ${\bar{\eta}} =  \left ( \frac{\displaystyle \int{\frac{1}{\eta} \text{d} x}}{\int{\text{d} x}} \right)^{-1} $.

So in 1D, the field can diffuse through the slab as quickly as it piles up, once the amplification factor has reached $R_M$. This provides a conceptual upper limit on the amplification of the magnetic field  as it piles up on the upwind side of the  asteroid.  

\subsection{Limits on amplification due to diffusion around the asteroid}\label{sec:td2}
In 3D, the magnetic field can diffuse around the asteroid as well as through - so if the diffusion of the wind material (turbulent, numerical, microphysical) is large enough, this can also limit the amplification to the corresponding Reynolds number. 

\begin{equation}
  A = R_{M\alpha} \equiv \frac{v_w R}{\alpha}
  \label{eq:rma}
\end{equation}
Here $ \alpha $ is the manually adjustable numerical diffusion present in the code, and is mostly dependent on the grid size.

\subsection{Limits on amplification due to pressure confinement}

We now  allow for transverse motion perpendicular to the wind flow away from the $\hat{x}$ axis. For the steady state wind flow exactly  along the $\hat{x}$ axis, we have from symmetry $v_y = v_z = B_x = B_z = 0$.  We can therefore write the mass continuity equation as  
\begin{equation}
    \frac{\uppartial }{\uppartial  x} \left [  \rho v_x \right]= \underbrace{-\rho \left (\frac{\uppartial  v_y}{\uppartial  y} + \frac{\uppartial  v_z}{\uppartial  z} \right)}_{\text{Transverse mass loss}}.
\end{equation}
This is  like the 1D case - but with an extra mass loss term involving the divergence of the transverse velocities.  Material can now be deflected along field lines in the $y$ direction or perpendicular to the field lines in the $z$ directions.

Correspondingly, the $\hat{x}$ component of the momentum equation reduces to
\begin{equation}
    \frac{\uppartial }{\uppartial  x} \left [ \rho v_x^2 + P + \frac{1}{2} B_y^2 \right ] = \underbrace{B_y \frac{\uppartial  B_x}{\uppartial  y}}_{\text{Magnetic tension}} \underbrace{- \rho v_x \left ( \frac{\uppartial  v_y}{\uppartial  y} +  \frac{\uppartial  v_z}{\uppartial  z} \right)}_{\text{Transverse mass loss}},
    \label{eq:pressure}
\end{equation}
which also has a corresponding transverse mass loss term acting to reduce the momentum flux approaching the body, along with a magnetic tension term that increases the momentum flux. 

The induction equation also has an additional term related to the transverse deflection of magnetic fields - but only in the $\hat{z}$ direction.  
\begin{equation}   
    \frac{\uppartial }{\uppartial  x} \left [ v_x B_y \right ]= \underbrace{ - B_y \frac{\uppartial  v_z}{\uppartial  z}}_{\text{Field line deflection in $\hat{z}$ only}}.
\end{equation}

These equations show that both mass, and the corresponding momentum and energy, can be redirected both along field lines (in $y$) and perpendicular to them (in $z$), while the magnetic field can only be redirected in the $z$ direction.

We can integrate the momentum equation towards the asteroid boundary to arrive at
\begin{align}
     \rho_b v_b^2 + P_b + \frac{1}{2} B_b^2 = &       \rho_0 v_0^2 + P_0 + \frac{1}{2} B_0^2 \nonumber \\ 
     &  + \displaystyle \int { \left[ B_y \frac{\uppartial  B_x}{\uppartial  y} - \rho v_x \left ( \frac{\uppartial  v_y}{\uppartial  y} +  \frac{\uppartial  v_z}{\uppartial  z} \right) \right] \text{d} x},
     \label{eq:momentumflux}
\end{align}
where $\rho_b$, $v_b$, $P_b$, and $B_b$ are the density, $x$ velocity, thermal pressure, and $y$ component of the magnetic field at the boundary with the asteroid and variabiles with subscript "0" are  the corresponding values of the wind. 

If we ignore the integral which involves the mass/field deflection and magnetic tension terms, we can rewrite  Equation (\ref{eq:momentumflux}) in terms of the magnetic amplification $A \equiv \frac{B_b}{B_0}$, incoming mach number $M$. and plasma $\beta$
\begin{equation}
    \left (1 - \frac{P_b}{P_0} \right) + \gamma M^2 \left(1 - A^{-1} \right)  + \frac{1}{\beta}\left(1 - A^2\right) = 0
\label{19}
\end{equation}
If we also assume that the movement of material along the field lines reduces the thermal pressure of the material at the asteroid surface so that $P_b \approx P_0$,  and  assume $A >> 1$, then Equation (\ref{19})  simplifies to 
\begin{equation}
    \gamma M^2 - \frac{A^2}{\beta} = 0 \rightarrow A = M \sqrt{\gamma \beta}.
\label{mrb20}
\end{equation}
This is equivalent to balancing magnetic pressure at the asteroid boundary with the incoming ram pressure of the wind and would be   an upper limit to the amplification when all other terms except magnetic tension are included.  Of all  the terms that we have dropped, only the magnetic tension term could increase the amplification beyond this value.  

Combining the ram pressure limit with the diffusion limits (both through and around the asteroid), we then predict the magnetic field amplification to be
\begin{equation}
    A \lesssim \max \left[ 1, \min \left( M \sqrt{\gamma \beta}, R_M, R_{M\alpha} \right) \right ]
    \label{eq:theory}
\end{equation}
This estimate is based on the underlying physics, but  in practice we must  carefully consider the influence of  numerical resolution and numerical diffusion when corroborating  Equation (\ref{eq:theory}) with what simulations actually show. We address this  further in \S \ref{sec:wind}.

\section{Results}
We carried out a suite of runs exploring the parameter space indicated in Table \ref{tab:summary}. Fig. \ref{fig:all} shows a plot of amplification vs magnetic Reynolds' number for all  cases. Fig. \ref{fig:xyAll} shows  the midplane slice of magnetic field, density, and velocity of a select few cases, while Fig. \ref{fig:lines} shows the 3D field distribution. The O and X type reconnection regions on the night side of the asteroid in Fig. \ref{fig:xyAll} are particularly notable. The insights we gain from these simulations are summarized for different cases.

\subsection{Summary of runs} 
Table \ref{tab:summary} lists our runs and their parameters along with  the peak measured field amplification, and the theoretically predicted value from Equation (\ref{eq:theory}).

\begin{figure}
	\centering
	\includegraphics[width=0.48 \textwidth]{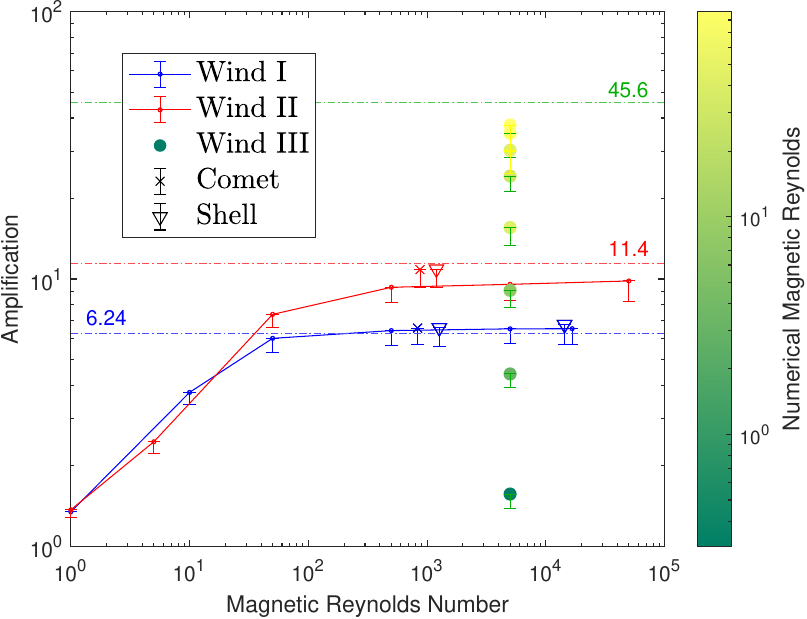}
	\caption{The maximum steady state magnetic amplification vs magnetic Reynolds number for simulations based on three different solar wind models (see Table \ref{tab:summary}). \textcolor{blue}{Wind I:} Ion density: 300 $\mbox{cm}^{-3}$, $ 500 \times 10^3 $ K, 500 $\text{km} \, \text{s}^{-1}$, $ M\sqrt{\gamma \beta} = 6.24 $; \textcolor{red}{Wind II:} Ion density: 1000 $\mbox{cm}^{-3}$, $ 500 \times 10^3 $ K, 500 $\text{km} \, \text{s}^{-1}$, $ M\sqrt{\gamma \beta} = 11.4 $; \textcolor[rgb]{ 0,  0.502,  0}{Wind III:} Ion density: 300 $\mbox{cm}^{-3}$, $ 150 \times 10^3 $ K, 500 $\text{km} \, \text{s}^{-1}$, $ M\sqrt{\gamma \beta} = 45.6 $. 	Solid curves show constant resistivity profiles having different resistive magnetic Reynolds numbers ($R_M$) but the same numerical magnetic Reynolds number ($R_{M\alpha} \sim 44$). Filled circles show the high $M\sqrt{\gamma \beta}$ case with fill color representing the value of $R_{M\alpha}$. Crosses show shell, and down triangles cometary cases. Dotted lines show theoretical  amplification for a given wind from Equation (\ref{mrb20}). High values of the error bars correspond to the global maxima at the dayside of the body directly perpendicular to the wind, while the low values are the L16 norm over a R/2 (250 km) region around it.}
	\label{fig:all}
\end{figure}

The amplification column refers to the peak magnetic field amplification, typically at the cell just outside the asteroid (including any outer shell  layer), and directly facing  the incident wind. There are other ways of computing the amplification, for example the $L^n$ norm where $n \geq 2$. In Fig. \ref{fig:all}, we add error bars to reflect this.

The theory column of the table lists the predicted amplification from a balance of the solar wind ram pressure with the magnetic field pressure, ignoring thermal pressure, provided we are above a threshold of conductive and numerical Magnetic Reynolds's number ($R_M$ and $R_{M\alpha}$) as per Eqn. \ref{eq:theory}. This somewhat overpredicts  the magnetic amplification for cases with higher $\beta$ (denser wind, in red in Figures), suggesting that thermal pressure contributes to the total pressure, thereby reducing the magnetic pressure needed to balance the ram pressure. 

\begin{table*}
	\centering
	\caption{Summary of Simulations (see text for  description of quantities listed)}
	\begin{tabular}{lr||lr|rrr}
		\toprule
		\multicolumn{2}{c||}{} & \multicolumn{2}{c|}{\textbf{Asteroid}} & \multirowcell{2}{$ \mathbf{R_{M\alpha}} $} & \multirowcell{2}{\textbf{Theoretical} \\ \textbf{Amplification}} 
		& \multirowcell{2}{\textbf{Simulated$^{**}$} \\ \textbf{Amplification}}\\
		
		\multicolumn{2}{c||}{} & \textbf{Structure} & $\mathbf{R_M^\dagger}$ &       &       &  \\
		
		\midrule
\multicolumn{2}{c||}{\textbf{Wind I}} & Shell & 1268  & 43.71 & 6.24  & 6.57 \\
Field (nT) & 100   & Shell & 14382 & 43.71 & 6.24  & 6.75 \\
Temperature (K) & $ 500 \times 10^3 $ & Comet & 832   & 43.71 & 6.24  & 6.55 \\
Ion density ($\mbox{cm}^{-3}$) & 300   & Fixed & 1     & 43.71 & 1.00  & 1.35 \\
Velocity ($\text{km} \, \text{s}^{-1}$) & 500   & Fixed & 10    & 43.71 & 6.24  & 3.77 \\
Mach  & 4.74  & Fixed & 50    & 43.71 & 6.24  & 6.00 \\
Beta  & 1.04  & Fixed & 500   & 43.71 & 6.24  & 6.41 \\
$ M \sqrt{\gamma \beta} $ & 6.24  & Fixed & 5000  & 43.71 & 6.24  & 6.51 \\
      &       & Fixed & 16667 & 43.71 & 6.24  & 6.52 \\
      \midrule
\multicolumn{2}{c||}{\textbf{Wind II}} & Shell & 1198  & 43.71 & 11.40 & 10.89 \\
Field (nT) & 100   & Comet & 872   & 43.71 & 11.40 & 10.85 \\
Temperature (K) & $ 500 \times 10^3 $ & Fixed & 1     & 43.71 & 1.00  & 1.37 \\
Ion density ($\mbox{cm}^{-3}$) & 1000  & Fixed & 5     & 43.71 & 5.00  & 2.46 \\
Velocity ($\text{km} \, \text{s}^{-1}$) & 500   & Fixed & 50    & 43.71 & 11.40 & 7.37 \\
Mach  & 4.74  & Fixed & 500   & 43.71 & 11.40 & 9.31 \\
Beta  & 3.47  & Fixed & 5000  & 43.71 & 11.40 & 9.55 \\
$ M \sqrt{\gamma \beta} $ & 11.40 & Fixed & 50000 & 43.71 & 11.40 & 9.83 \\
\midrule
\multicolumn{2}{c||}{\textbf{Wind III}} & \textit{Fixed*} & \textit{5000} & \textit{87.43} & \textit{45.60} & \textit{37.64} \\
Field (nT) & 14    & Fixed & 5000  & 87.43 & 45.60 & 35.08 \\
Temperature (K) & $ 150 \times 10^3 $ & Fixed & 5000  & 43.71 & 43.71 & 15.58 \\
Ion density ($\mbox{cm}^{-3}$) & 300   & \textit{Fixed*} & \textit{5000} & \textit{43.71} & \textit{43.71} & \textit{30.29} \\
Velocity ($\text{km} \, \text{s}^{-1}$) & 500   & Fixed & 5000  & 30.60 & 30.60 & 24.26 \\
Mach  & 8.65  & Fixed & 5000  & 7.65  & 7.65  & 9.07 \\
Beta  & 16.66 & Fixed & 5000  & \textit{3.06} & 3.06  & 4.42 \\
$ M \sqrt{\gamma \beta} $ & 45.60 & Fixed & 5000  & 0.31  & 1.00  & 1.57 \\
		
		\bottomrule
		\multicolumn{7}{l}{*Wind III \textit{Italicized} cases have double grid resolution.} \\
		\multicolumn{7}{l}{$^\dagger \ R_M $ values for Shell and Comet cases are weight averaged over current.}\\ 
		\multicolumn{7}{l}{$^{**}$ \ Values represent  the  { peak field} of all grid cells. See Fig. \ref{fig:all} for error bars.}
		\\
		
	\end{tabular}%
	\label{tab:summary}%
\end{table*}%

We computationally demonstrate that early solar wind conditions could have caused sufficient pileup to induce the  NRM observed in some carbonaceous chondrite meteorites \citep{obrein20}. We found pileup to be close to our theoretical estimate of $ M \sqrt{ \gamma \beta} $, with the dependence on $ \beta $ being generally slightly weaker, as thermal pressure also contributes to the pressure equilibrium) for sufficiently conductive asteroid surfaces (see Fig. \ref{fig:all}). This field pileup saturates at a threshold conductivity above which the magnetic and thermal pressure balance the incoming solar wind ram pressure. The timescale to reach the steady state is proportional to a body-shell averaged diffusion timescale. 

We find that both the value of the effective conductivity and whether the conductivity is uniform or dominated in a shell both influence the amount of  magnetic field pileup.  Profiles having an outer conductive $10\%$ shell, as in our cometary and shell models, have higher pileups than cases with same effective conductivity distributed uniformly. However, this field is  at the outer boundary of the shell. If  this shell represents an atmosphere rather than a body surface layer, turbulent mixing of atmospheric layers would be required for the field to reach the main body.

\subsection{Constant profile cases} \label{sec:const}
Most of our simulations in Table \ref{tab:summary} use a constant resistivity profile. As Fig. \ref{fig:all} shows, the simulated amplification increases almost linearly with $ R_M $, and then asymptotically approaches the theoretical value  from Equation (\ref{mrb20}). The vertical points of different shades of green in the figure show how $R_{M\alpha} $ affects the amplification for a wind with very high $\beta$. There, for a fixed $R_M$, as $R_{M\alpha} $ increases the saturation value again approaches Equation (\ref{mrb20}) and  validates Equation (\ref{eq:theory}).

\begin{figure*} 
    \centering
    \includegraphics[width=0.89 \textwidth]{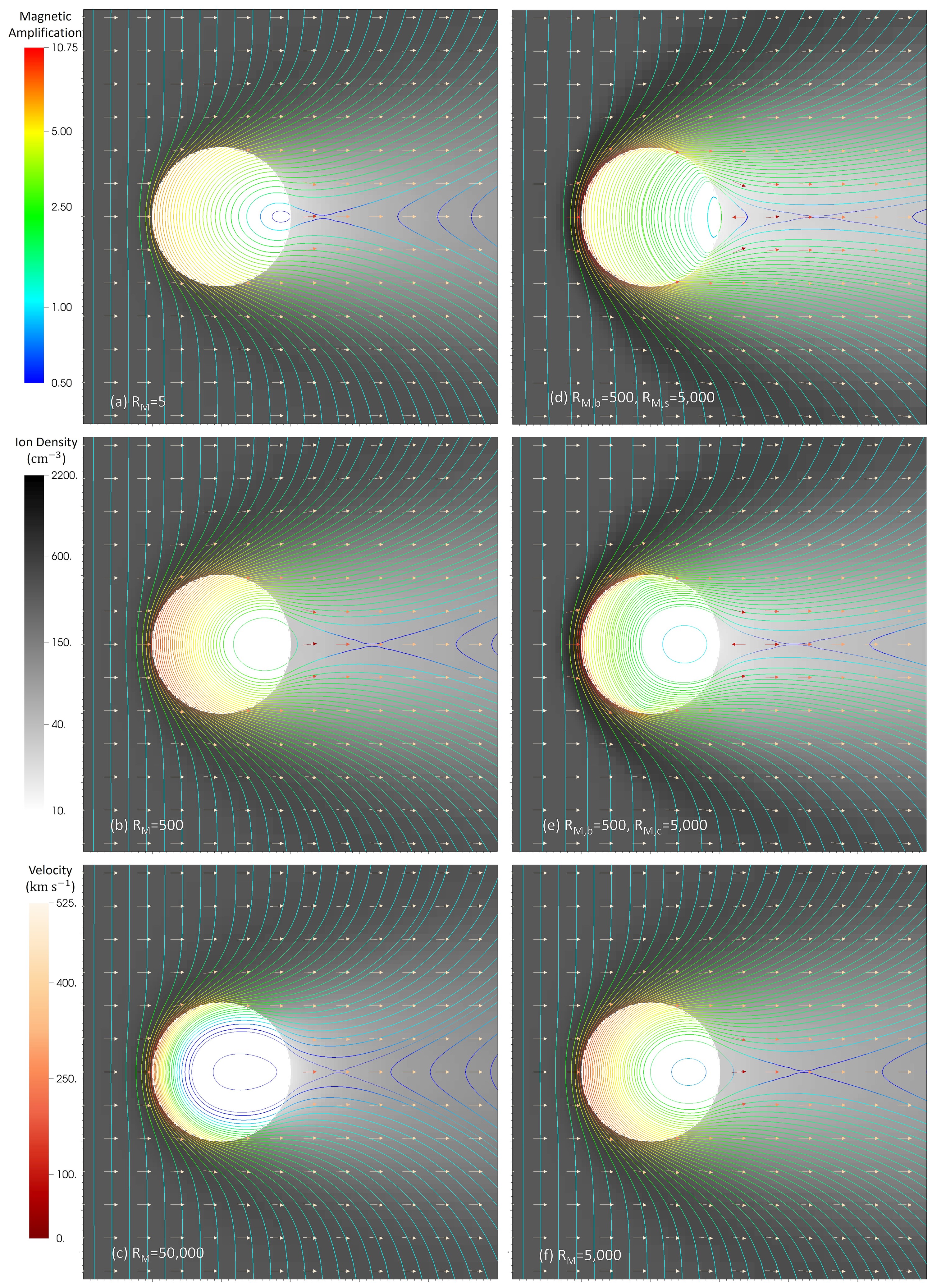}
    \caption{Steady state Magnetic amplification, particle density, and velocity in the xy mid-plane for select cases with $ T_w = 500 \times 10^3 K $, $ \rho_{i,w} = 1000 \mbox{cm}^{-3} $, $ v_w = 500 km s^{-1}$. The velocity scale is linear while others are logscale. {\bf Left column panels}:  show progression in uniform resistivity cases with no shell layer, (a) Top: $ R_M = 5 $; (b) Middle: $ R_M = 500 $; (c) Bottom: $ R_M = $ 50,000. 
    {\bf Right column panels:} (see also Fig. \ref{fig:lines})
    (d) Top: Body with $ R_M = 500 $, $10 \%$ full outer shell $ R_M = $ 5,000; (e) Middle: Body with $ R_M = 500 $, $10 \%$ Cometary dayside shell $ R_M = $ 5,000; (f) Bottom: Constant resistivity profile with $ R_M = $ 5,000. }
    \label{fig:xyAll}
\end{figure*}

Figure \ref{fig:xyAll} shows the steady state magnetic amplification, density, and velocity for the denser ($ 1000 \mbox{cm}^{-3} $) wind for $ R_M = 5 \times 10^4, \ 5000, \ 500 $, and $ 5 $. As expected, the density, and magnetic field pileup increase with the magnetic Reynolds number while the wind velocity remains nearly the same. The lightening in shade on the night side is indicative of the rarefaction there. In general, the mass pileup is much smaller than the field pileup as particles can easily leave in directions parallel to the field and perpendicular to the incoming wind near the body.

\begin{figure}
	\centering
	\includegraphics[width=0.48 \textwidth]{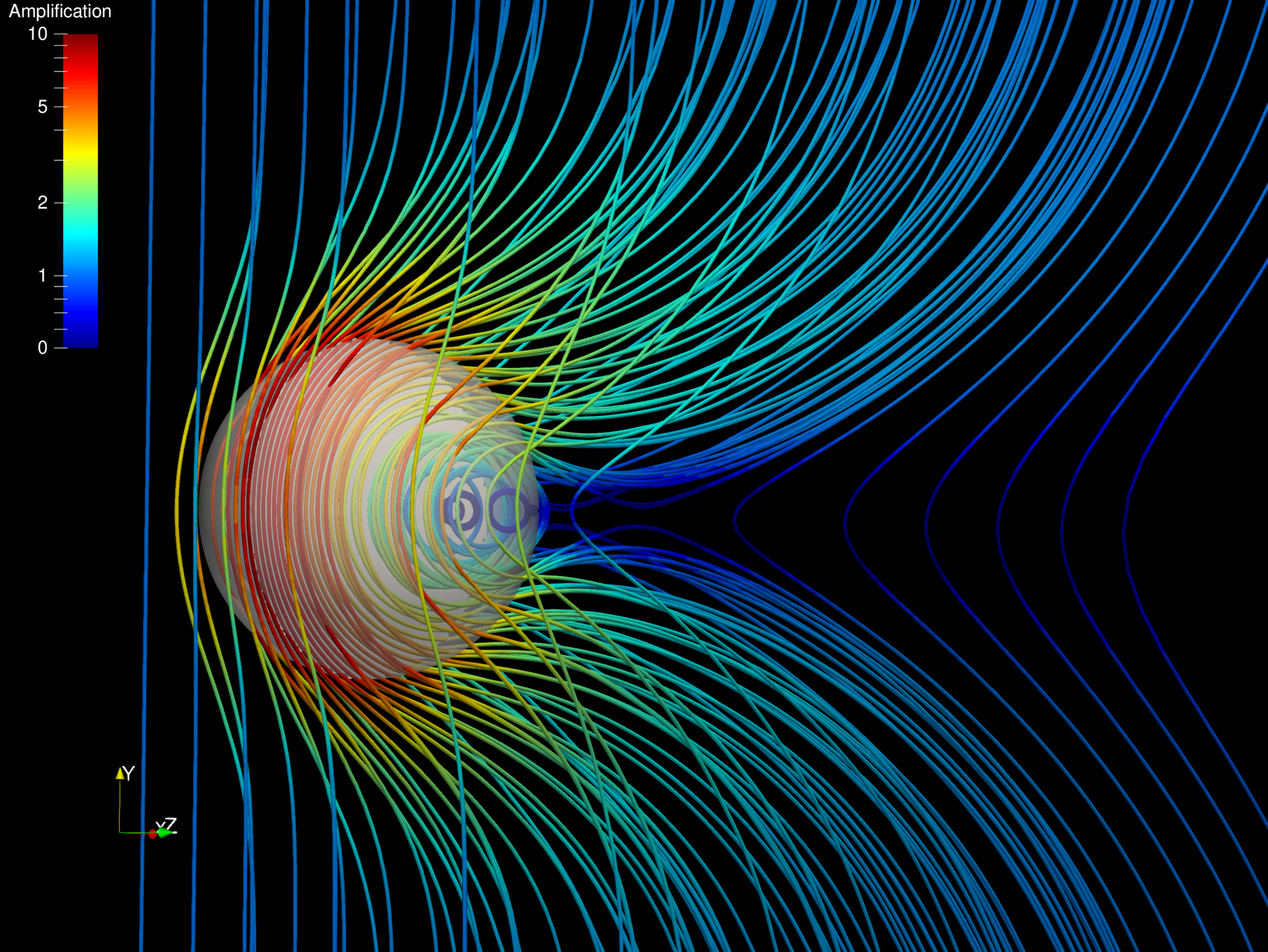}
	\caption{(\href{https://skfb.ly/o6sKt}{Link to 3D Model}) Steady state view of magnetic field lines draping over the surface of the asteroid (shown in translucent white). The field lines are colored by their strength relative to the incoming solar wind values on a log scale (see PDF placeholder image for color bar) from 0.5 (blue) to 10 (red). The field lines were seeded from points evenly spaced along a line across the bottom boundary ($y = -12 R_{\text{asteroid}}$, $z = 0$), as well as points within a sphere of radius $1.2 R_{\text{asteroid}}$ centered on the asteroid. This plot corresponds to our denser solar wind case ($ T_w = 500 \times 10^3 K $, $ \rho_{i,w} = 1000 \mbox{cm}^{-3} $, $ v_w = 500 km s^{-1}$), and a constant $ R_M = $ 5,000 inside the asteroid.}
	\label{fig:model_streamlines}
\end{figure}

Fig. \ref{fig:model_streamlines}, shows a \href{https://skfb.ly/o6sKt}{3D model}\footnote{Link to 3D Model in Fig. \ref{fig:model_streamlines}: https://bit.ly/anand2021} of the final field configuration for the constant resistivity profile case with $R_M=$ 5,000.

\subsection{Shell case}
In this case the outer $ 10\% $ of the shell in radius is 10 times more conductive than the main body. We ran 3 distinct cases of this kind: two with an asteroid body resistivity of $ 1250 \, \Omega \text{m} \ \text{or} \ 0.2 \% $ of the critical resistivity (see \S \ref{sec:res} for definition) but different wind models; and a 10x more conductive case for the $ 300 \mbox{cm}^{-3} $ wind value. These cases are listed in the first three rows of Table \ref{tab:summary}. These cases correspond physically to a body with a thick outer conducting shell or atmosphere. In principle, we could  further reduce  the conductive layer thickness down to the resolution of the grid ($\sim 1.6 \%$ of asteroid).

Fig. \ref{fig:xyAll}, (d) top right panel, and Fig. \ref{fig:lines} (a) top panels show the steady state frame for the shell case for the denser wind. We find the peak amplification of this case (and the cometary case discussed below) to be higher than all constant profile cases, irrespective of $ R_M $, even though the conducting shell itself has a $ R_M = $ 5,000. But the field distribution in Fig. \ref{fig:xyAll} (top-right panel) shows that the peak is more localized within the shell as compared to its broader distribution for the constant profile cases.  From the RHS term in Equation (\ref{eq:Eu4}) and the tension term in Equation (\ref{eq:momentumflux}), we do expect a sharper gradient and higher field inside the shell. 

In Fig. \ref{fig:all} and  Table \ref{tab:summary} we note that the  shell cases for Wind I overshoot the prediction of theoretical maximum from Equation (\ref{eq:theory}), however, this is well within our error bars and not a cause of signficant disagreement between theory and simulation. Future work should explore the parameter space imposed by the shell thickness and conductivity.

\subsection{Cometary case} 
The asteroid body in this case has a resistivity of $ 2 \times 10^{-3} \eta_c $ or $R_M = 500 $, with only the dayside of the $ 10\% $ shell having $R_M =$ 5,000. We ran a simulation for both wind cases. These runs result in higher magnetic field amplification than any of the constant profile cases, but lower than the isotropic shell cases. Physically, this case  represents bodies outgassing volatiles from the radiation and incident wind from the host star.  
\footnote{Even without the shell, we  will show that amplifications  $ > 9 $ occur \textit{inside} the body, which is sufficient to explain experimental measurements of fields from the select CV and CM meteorites highlighted by \cite{obrein20}. We shall see that  the shelled cases allow for even more field pile up. However, when this shell represents an atmosphere,  the body beneath could  be partially shielded from field pileup in absence of  turbulent mixing. }

\begin{figure*}
	\centering
	\includegraphics[width=0.99 \textwidth]{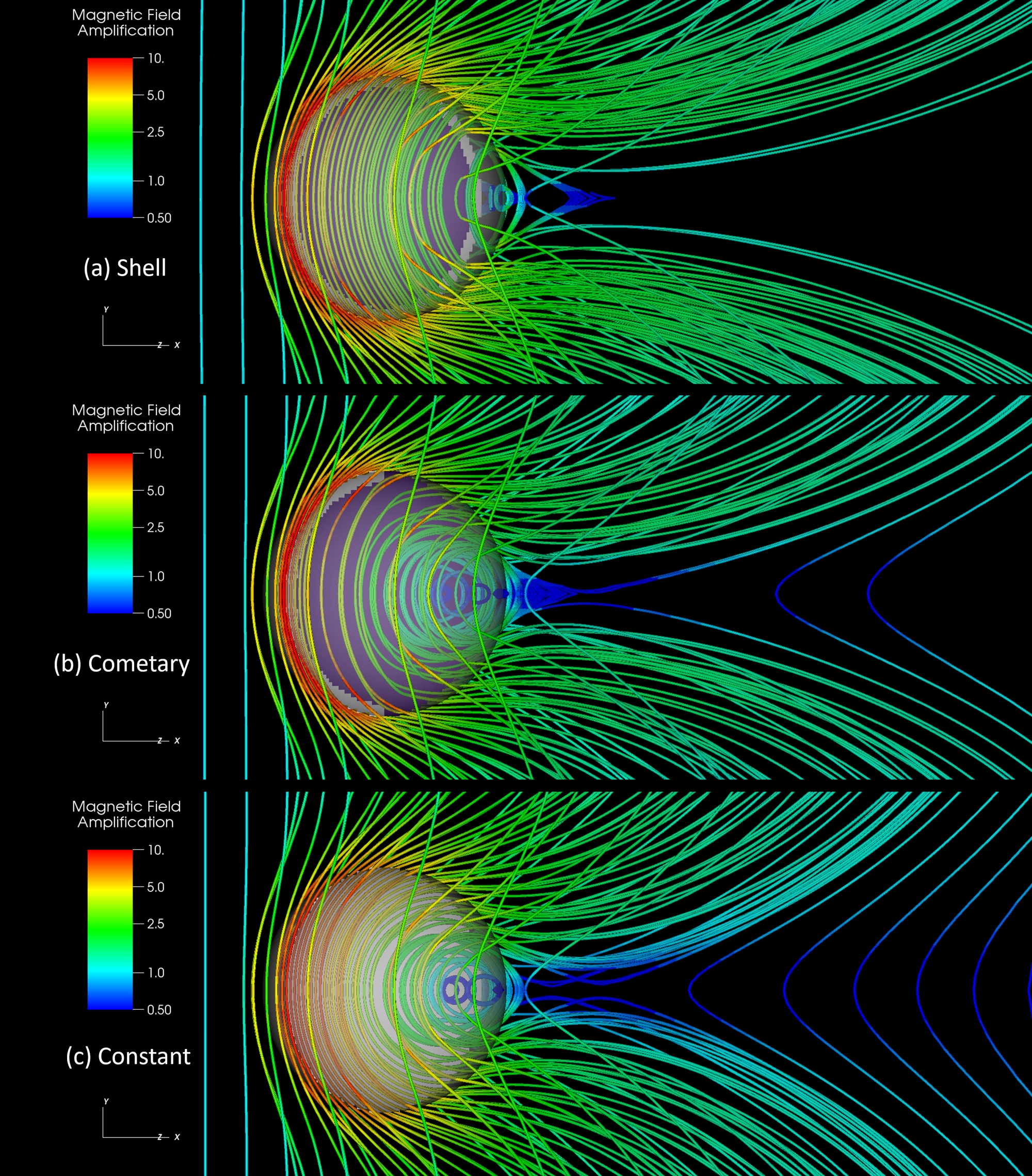}
	\caption{3D Magnetic Field Structure at quasi-steady state for a subset of cases corresponding to the right column of Fig. \ref{fig:all} at steady state. Shaded violet region in the xy plane inside the body represents a $R_M=500$, and greyish-white regions $ R_M = $ 5,000. Cases (a) {\bf Top}: Body with $ R_M = 500 $, $10 \%$ full outer shell $ R_M = $ 5,000;  (b) {\bf Middle}: Body with $ R_M = 500 $, $10 \%$ Cometary dayside shell $ R_M = $ 5,000; (c) {\bf Bottom}: Constant resistivity profile with $ R_M = $ 5,000. The strength of amplification increases substantially with the presence of a conducting shell.}
	\label{fig:lines}
\end{figure*}

Fig. \ref{fig:xyAll}, (e) middle right panel and Fig. \ref{fig:lines} (b) middle panel show the steady state field configuration for the denser wind  run of the comet cases. The cometary case differs from the full shell case   in that  the  field coming in from the back side is smoothly varying (\textit{i.e.} has no kinks). For the shell case, the field can have a kink at the boundary--allowing  higher diffusion inside the body which provides a more uniform field therein. We also reported on this case in \cite{obrein20}, Fig. 2 and a video in the supplementary material. This contrasts the 3D model (Fig. \ref{fig:model_streamlines}) of the constant resistivity profile case for $R_M=$ 5,000 at the end of \S \ref{sec:const}. As in all cases, the peak field is in the zone just outside the boundary on the dayside of the asteroid. 
Fig. \ref{fig:all} and Table \ref{tab:summary} show that the comet case for Wind I (like the shell case above) overshoots the prediction of the theoretical maximum from Equation \ref{eq:theory}, but this is again well within our error bars and again not a significant  disagreement between theory and simulation.

\section{Interpretation, applicability, and limitations}

\subsection{Role of solar wind} \label{sec:wind}
The ram, thermal, and magnetic energy of the solar wind supply the source of  magnetic energy at the dayside boundary of the parent body. If the conductivity of the parent body surface is high enough to sufficiently stall the flow of solar wind, the magnetic field at the boundary keeps increasing until this magnetic pressure can balance the ram, thermal, and magnetic energy of the solar wind. Figure \ref{fig:all}  and Table \ref{tab:summary} show this trend.  Note that the the maximum $A$ in all cases would be better estimated from solving Equation (\ref{19}) than from the limiting case of Equation (\ref{mrb20}) but, but for simplicity we discuss the results in terms of how close they approach the simple limit
of the latter.

The blue curve, having moderate solar wind parameters of $300 \mbox{cm}^{-3} $, $5 \times 10^5$ K, 100 nT is indeed  nearly solar wind limited and the conversion from solar wind pressure to surface magnetic pressure happens relatively efficiently and the value of Equation (\ref{mrb20}) is approached.
The red curve had a higher density, hence the mass loss, and ram pressure terms on the left side of Equation (\ref{eq:momentumflux}) ultimately contribute, lowering the values of the magnetic pressure needed to balance the incoming pressure.  Thus the simulated values of $A$ are lower than the red line of $ M \sqrt{\gamma \beta}$ from Equation (\ref{mrb20}) which assumes only magnetic pressure balances the incoming wind pressure.
The green curve has a very high $\beta$, and hence even lower relative contributions from the magnetic pressure term are needed to balance the incoming pressure and so again slightly lower values achieved than the value of Equation (\ref{mrb20}).

Current theory and observations of systems analogous to our solar system suggest that the red wind case ($10^3 \mbox{cm}^{-3} $, $ 5 \times 10^5$ K, 100 nT) would be closest to the values expected for our sun.

\subsection{Effect of numerical resistivity} \label{sec:diffAlpha}
In Fig. \ref{fig:all}, the green-yellow circles show the effect of the imposed numerical diffusion  into of our code. The values of the red and blue curves from Equation (\ref{eq:theory}) are ultimately limited by the solar wind. The points on the left region of Fig. \ref{fig:all} ($ R_M < 50 $) are limited by the magnetic Reynolds number of their true resistivity. However, the resolution of the code introduces a numerical diffusion magnetic Reynolds number ($R_{M\alpha}$) which we defined similarly to $ R_M $ in Equation (\ref{eq:rma}). 
 
\begin{figure}
	\centering
	\includegraphics[width=0.48 \textwidth]{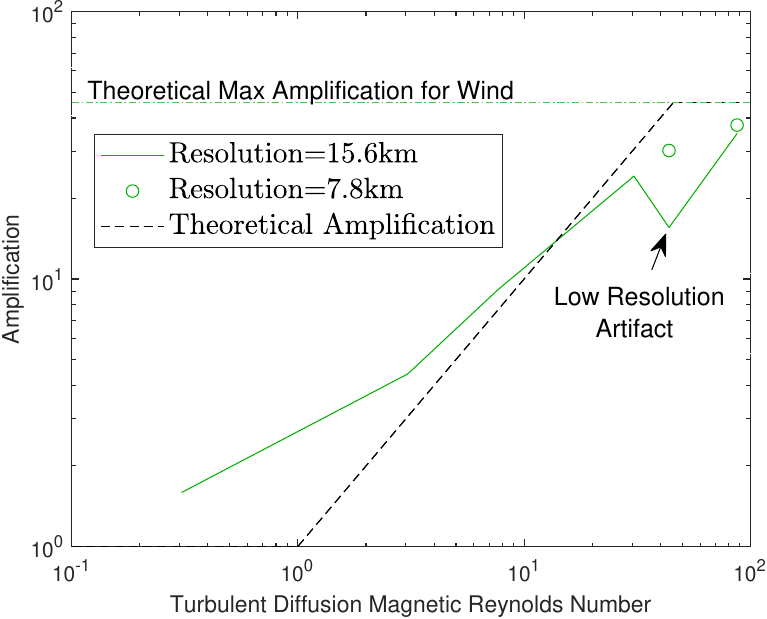}
	\caption{Amplification vs numerical magnetic Reynolds number for the Mach=8.65, Beta=16.66 wind.  The horizontal dashed green line shows the value from Equation (\ref{mrb20}).  The rising black dashed lines shows the theoretical  prediction for $A$ from  Equation (\ref{eq:theory}) which increases with $R_{M\alpha}$ until $ R_{M\alpha}> Min (R_M,M \sqrt{\gamma \beta})=\sqrt{\gamma \beta}$. The solid green line shows the results from simulations for cases with lower resolution, while points show cases with double the resolution to highlight the fact that the dip in the solid green curve is a numerical artifact. Overall, the simulations follow the theoretical trend of the prediction.}
	\label{fig:da2_b14}
\end{figure}

For the blue and red curves,  $ R_{M\alpha} $  is a few times higher than the limitation imposed by the stellar wind pressure so we do not see the effect of numerical diffusivity. In contrast,  the green wind model has a very high beta (16.66) and high mach number (8.65), i.e. wind parameters are 300 $ \mbox{cm}^{-3} $, $1.5 \times 10^ 5$ K, 500 $\text{km s}^{-1}$, 14 nT and so $A$ is thus limited  by $R_{M,\alpha}$, as expected from Equation (\ref{eq:theory}).  For this case, Fig. \ref{fig:da2_b14} shows how the amplification changes with $R_{M,\alpha}$ for  2 different grid resolutions (label shows length of smallest cell). The dip  towards the upper right  is a numerical artifact of a low resolution grid. This highlights  that the grid size chosen must resolve the physics near the surface of the body. 

\subsection{General applicability}
The preceding analyses can readily be applied to any asteroid bodies having a constant resistivity profile as seen in Fig. \ref{fig:all}. 
The shell and cometary profiles are applicable to   asteroids having a very conductive body. We have only used profiles where the outer $ 10\% $ shell has $10\%$ the resistivity of the body, but we expect the amplification to hold for thicker shells. For much thinner shells the results may be  particularly affected if the shell is so thin that  the diffusion time across the shell becomes comparable to or less than the flow crossing time. 
The result  would then be equivalent to a constant profile case with the shell playing little role.  If the thicker shells represent atmospheres, then turbulent mixing would be needed for the asteroid body magnetization to gain substantially from the increased pileup.
Importantly, the needed pile up to explain  some paleofields, as determined  by studies of meteorites, is not contingent upon having a shell layer as our constant profile cases have amplifications within $10\%$ of the cometary and shell case values.

In short, our results suggest that apart from the most resistive objects, all bodies considered are able to pile up magnetic fields. The effect is  most pronounced for bodies with a 
 conductive shell.

{The ambient magnetic fields from our WIM MHD models can explain why some carbonaceous meteorites with reliable magnetic mineral recorders formed after the dispersal of nebular gas and dust record magnetizations, even though their parent bodies where undifferentiated and thus lacked core dynamos \citep{obrein20}. The presence of an ambient magnetic field produced by the WIM mechanism may explain the magnetization of other meteorites, but as noted earlier, reliable magnetic mineral carriers must be present able to record and preserve ambient field records, and geochronological data should constrain formation ages to after the dispersal of the protoplanetary gas and dust. Rapid magnetization acquisition times can assist in the preservation of unidirectional magnetizations \citep{obrein20}. However, this is not necessarily required for coherent magnetizations to be imparted to near surface rocks on asteroids. Instead, this will be a function of the distribution of blocking temperatures or blocking volumes in a given magnetic mineral assemblage \citep[see][, in particular Supplementary Information Sections 4, 8]{obrein20}.}

{In the case where reliable magnetic mineral recorders are present, and their ages are known to postdate dispersal of nebular gas and dust, the paleointensity of a meteorite together with ambient fields from our WIM models can also be used to constrain the orbital distance of a parent body at the time of magnetization, providing important constraints on Solar System evolution. In this way, \cite{obrein20} were able to show that paleointensity data and WIM models indicated arrival of the CV and CM parent bodies from the outer Solar System to the heliocentric distance of the present-day asteroid belt by 5 million years after the formation of CAIs.

In addition, the WIM mechanism may explain magnetizations observed in future space missions to asteroids, such as (16) Psyche, where the nature of the body - in the case of Psyche the origin of its metallic surface - is presently unknown \citep{Elkins-Tanton+2020,Johnson+2020} and therefore the present/absence of a past core dynamo is similarly undetermined.

\subsection{Implications for interpreting previous work}
Although the work of \cite{obrein20}, supported the  possibility originally discussed in \cite{tarduno2017lpi} that  WIM
might explain the  the magnetization of undifferentiated  meteorite parent bodies,  \cite{oran18} produced a series of magnetohydrodynamic (MHD) simulations that were used to  argue that the solar wind was an insufficient source. In fact there is no contradiction between the simulations results in their paper and ours here.

\cite{oran18} used a  BATS-R-US simulation with  a different solar wind model and body resistivity type. While they found amplifications of only about $ 3 $, this is perfectly in line with the maximum amplification that our theory predicts for a $ \mathrm{35 \, \mbox{cm}^{-3}, \ 700 \, km \, s^{-1}, \ 50,000 \, K} $ solar wind used in their simulations. In fact we ran a case using exactly their wind model and resistivity and found a steady state amplification values equal to their to within a few percent. This  verifies that both simulation methods are mutually consistent and that there is no contradiction between their computation and ours.

The model parameter choice is not the only source of the difference in conclusions reached by \cite{obrein20} and those of \cite{oran18}. Specifically, \cite{oran18} also sought to explain the high ($\mathcal{O} \sim $ 10 $\mu$T) apparent paleointensity value from the Allende meteorite \citep{Carporzen2011, weiss2013}, and they averaged solar wind magnetizations across many reversals. As noted earlier, \cite{obrein20} showed that the Allende magnetization is not a reliable paleointensity recorder, whereas meteorites with reliable recording properties yield paleofield strengths an order of magnitude less than the now invalid Allende value. Moreover, the averaging of solar wind magnetizations over many reversals is incompatible with the unblocking temperature 
data and evidence for a short duration of magnetic mineral formation summarized 
by \cite{obrein20}. Nevertheless, \cite{weiss21} cited \cite{oran18} to claim that the solar wind was “more than 2 orders of magnitude too weak to explain CM magnetizations”. The results of \cite{obrein20} and our expanded modeling here show that this claim is unjustified.
In particular, here we move beyond \cite{obrein20} to show more generally how amplifications much higher than those modeled by \cite{oran18} are possible with different sets of parameters that are realistic for the early solar wind.

\subsection{Limitations and approximations}

Here we comment on how some of the limitations and approximations affect the results:
\begin{itemize}
	\item \textbf{Cartesian Grid:} AstroBEAR currently does not have an option for running simulations in spherical coordinates. Hence, there are small ``staircase'' like artifacts on the surface of the asteroid in our models. These features decrease  with resolution (or AMR level used), but the tiny increase in accuracy was not worth the 
	increase in memory and core hours. For   cases with low resistivity, some cells on the dayside 
	oversaturated with magnetic field, causing small time steps and  slowing the simulation to a halt. We circumvented this problem by using a controlled numerical diffusivity, characterized by our $R_{M_\alpha}$, for all flows outside the asteroid. This sped up the steady state convergence efficiently, with 256 core hours being sufficient for low to moderate resistivity.
	
	\item \textbf{Rarefactions on the Night Side:} As Fig. \ref{fig:xyAll} shows, the night side of the asteroid has the lowest densities in the plot region. In several cases this required use of  \textsc{AstroBEAR}'s minimum density protections to avoid the extremely high velocites that would otherwise  cause the time steps to catastrophically decrease and halt the simulation. 
\end{itemize}

Several other approximations  warrant mention:
\begin{itemize}
	\item \textbf{Size of Parent Body:} We  have taken the size to be 500 km. Our use of an MHD code  is necessarily incomplete given that the wind is really a collisionless plasma, but the approximation is not unreasonable  as long as the  body size is much larger than the gyroradius of the ions.
	
	\item \textbf{Revolution of the Asteroid:} We calculated the axial component of the Tidal force ($ \boldsymbol{a}_{t, \, \text{axial}} \sim \Delta \boldsymbol{r} G M_\odot/R^3 \sim 5 \mathrm{nm \, s}^{-2}  $), and the Coriolis Force ($ \boldsymbol{a}_c = 2 \boldsymbol{\Omega} \times \boldsymbol{v'} \sim 1.48 \mathrm{mm \, s}^{-2} $) assuming the asteroid to be at 2 AU, and found that these can be neglected. Hence, we do not benefit from using a frame of reference co-rotating with the asteroid.
	
	\item \textbf{Magnetic permeability of Asteroids:} Here we have not accounted for the magnetic permeability of the asteroids. We expect it to be the same order of magnitude as vacuum, but including it could increase the field strength actually experienced by the body by $\gtrsim 10\%$  depending on the material.
\end{itemize}

\section{Conclusion}

From computational simulations and theoretical analyses,  we have shown that magnetic flux pile up from an impinging stellar wind can produce a substantial ambient magnetic field on an asteroid surface. As applied to the early solar wind, which would have had a higher ram pressure than that of the present day, we find that the ambient solar find field can be amplified by factors $A>9$ even without a shell of enhanced conductivity above that of the main body. 
This can explain the magnetizations found in some meteorites having reliable magnetic mineral records formed after dispersal of the protoplanetary disk, but that are thought to have come from undifferentiated parent bodies lacking internal (core dynamo) magnetic fields.

With a conductive shell layer  surrounding a parent body, we find that even larger amplification factors, $ A > 10 $, in the shell layer are likely} for  early solar wind properties well within  theoretical model limits and observational  constraints. The solar wind properties, and the effective resistivity of the asteroid's shell layer  emerge as the two most important factors affecting the amount of magnetic field pileup.
  
The WIM mechansim and the predicted amplification scaling are quite general, and can be applied to any magnetized stellar wind and asteroid combination. The WIM mechanism is also potentially important to  explain magnetizations of asteroids measured from future space missions such as (16) Psyche.
The results would ultimately be applicable to extrasolar planetary systems as well. 

Finally, we  note there are important microphysical aspects of the wind-asteroid interaction that we have not included, and  warrant  higher fidelity modeling the future.  At the MHD level this could include anisotropic viscosity due to the magnetic field which may be important at the wind-asteroid interface. If our shell cases were interpreted to model bodies with atmospheres, they should ultimately be augmented to include atmospheric flows which could lead to mixing within the atmosphere and/or between the atmosphere and surface layer. This  will be important for   determining how much of the piled up field at the top of an atmosphere could reach the  solid surface layers.
In addition, the microphysics of how the field is separated from plasma and  deposited into solid asteroid material at the interface is not directly treated in the MHD approximation and could benefit from computational models that that follow particles below at ion-gyroradii scales and/or have plasma-solid interaction capability. 
There is substantial opportunity for further work in all of these  directions.

\section*{Acknowledgements}
We thank the editor and reviewer for helpful comments that improved the manuscript.
We thank A.N. Krot for discussions.
AA acknowledges a Horton Graduate Fellowship from the UR Laboratory for Laser Energetics. The authors thank the Center for Integrated Research Computing (CIRC) at the University of Rochester for providing computational resources and technical support. Support provided from grants  NSF PHY-2020249, NASA grant 80NSSC19K0510, NSF EAR-2051550.

\section*{Data Availability}
The data underlying this article will be shared on reasonable request to the corresponding author.

\bibliographystyle{mnras}
\bibliography{bibliography.bbl}

\bsp	
\label{lastpage}
\end{document}